\documentclass{article}%
\usepackage{amsmath}
\usepackage{amsfonts}
\usepackage{amssymb}
\usepackage{graphicx}%
\usepackage{hyper}
\providecommand{\U}[1]{\protect \rule{.1in}{.1in}}

\begin{document}

\title{\Huge Tunnelling through black rings}
\author{Liu Zhao\thanks{\em email: lzhao@nankai.edu.cn}
\\Department of Physics, Nankai University,\\ Tianjin 300071, P R China}
\date{}
\maketitle


\begin{abstract}
Hawking radiation of black ring solutions to $5$-dimensional Einstein-Maxwell-dilaton gravity
theory is analyzed by use of the Parikh-Wilczek tunnelling method. To get the
correct tunnelling amplitude and emission rate, we adopted and developed the Angheben-Nadalini-Vanzo-Zerbini
covariant approach to cover the effects of rotation and
electronic discharge all at once, and the effect of back reaction is also taken into account.
This constitute a unified approach to the tunnelling problem. Provided the first law of thermodynamics for
black rings holds, the emission rate is proportional
to the exponential of the change of Bekenstein-Hawking entropy. Explicit calculation for black ring temperatures
agree exactly with the results obtained via the classical surface gravity method and the quasilocal formalism.
\end{abstract}

\section{Introduction}

The study of black hole radiation has long been an attractive field of study
in gravitational physics as well as in modern theoretical frameworks such as
string/brane dynamics. The reason for the problem of black hole radiation
keeps attracting attentions of theoretical physicists is partly due to the fact
that it indicates black holes are highly excited quantum states and therefore
can only be expected to be fully understood in terms of quantum gravity; on the other hand, the
understanding of black hole radiation is the
key to make the second law of thermodynamics in spacetimes involving black holes consistent.
Traditional approaches for the black hole
radiation involve studying of QFT in a (fixed) curved spacetime in which a
gravitational collapse occurs which lead to a purely thermo spectrum \cite{Hawking:1974sw}.
A more recent recurrence of interests on the black hole radiation problem has appeared,
with emphasis on the inclusion of back reaction or gravitational self-interaction of the emitted particles
on the black hole geometry \cite{gr-qc/9408003, hep-th/9411219, hep-th/9610045}, which result in a non-thermo rediation
spectrum. In particular, in \cite{hep-th/9610045}, Keski-Vakkuri and Kraus obtained a remarkable
relationship between the emission rate and the entropy loss of the black hole, which is interpreted
as the effect of the particle's tunnelling across the horizon. Such a relationship, and together with the
corresponding non-thermo spectrum, are believed to be helpful in understanding the information loss paradox
and the underlying unitarity problem. The tunnelling method is then
purified and popularized by the work of Parikh and Wilczek \cite{hep-th/9907001, hep-th/0204107, hep-th/0405160},
which is now known in the literature as Parikh-Wilczek tunnelling method.
A lot of works has already been carried out for further development of this
approach \cite{hep-th/0012134, hep-th/0108147, hep-th/0109108, hep-th/0110289,
hep-th/0207247, hep-th/0209185, hep-th/0503081, gr-qc/0504113,
gr-qc/0505015, Zhang:2005xt, hep-th/0511123, hep-th/0511250, gr-qc/0512153},
most of them being focused on the study of Hawking radiation
process from various black hole spacetimes and a few exceptional examples are
\cite{hep-th/0512121, gr-qc/0601018} which discussed the consistency of the tunnelling approach with the
classical laws of thermodynamics, \cite{hep-th/0505266}, which considered the Plank scale corrections
to the radiation spectrum via tunnelling method, and \cite{hep-th/0309092, hep-th/0405186} which studied
the corrections to the Cardy-Verlinder entropy due to the tunnelling effect.

In this work, we shall concentrate ourselves to the problem of Hawking
radiation of a certain class of black ring spacetimes. Unlike the usual black
hole spacetimes in $4$-dimensions which are found almost right after the birth
of Einstein's general theory of relativity, black ring solutions are
explicitly constructed very recently. One reason that such spacetimes were not
found until recently is that they are purely higher dimensional objects which
have no $4$-dimensional analogues. The topology of their event horizons is
also quite exceptional ($S^{2}\times S^{1}$, as their name indicates), and
perhaps the most striking feature that black ring solutions brought to us is
their infinite non-uniqueness, i.e. there exists infinitely many different
black ring solutions carrying the same mass, angular momentum and/or
electronic charge. In short words, black rings are the first explicit example
for the non-uniqueness of black hole spacetimes in higher dimensions.
Moreover, it was found in \cite{hep-th/0402149} that black ring spacetimes are actually some
very special intersecting brane configurations in string theory frameworks,
and as such they provide us with a new class of theoretical laboratories for
testing the string theory methods in counting the microstates of black
holes/rings. To this end we feel that is also interesting to study various
properties of black ring spacetimes -- including the thermodynamic properties
and radiation properties -- using the more traditional methods, with the hope
to make cross tests with  results from string theory and/or other approaches.

The paper is organized as follows. In section 2 we give a brief description of
the black ring solutions we shall study and extract some of their common
properties in order to make the forthcoming analysis in a more unified form.
In section 3, we first describe the tunnelling method adapted for the black
ring solutions, which is basically the method of Angheben,
Nadalini, Vanzo and Zerbini (henceforth referred to as the ANVZ approach)
\cite{hep-th/0503081, hep-th/0511250} for rotating black holes
with modifications to include the effect of electronic discharge, and the
effect of back reaction is also included according to \cite{gr-qc/0504113}. The explicit
results for the black ring spacetimes outlined in section 2 is then given
after the presentation of the general framework, and some discussion on the Kaluza-Klein cases
are also presented. Section 4 contains some
discussions and concluding remarks.

\section{Black ring solutions of Einstein-Maxwell-dilaton gravity}

A significant amount of black ring solutions have been found since the
pioneering works \cite{hep-th/0110258, hep-th/0110260}. The black ring spacetimes we shall study are only a
special subset of them, i.e. those which are solutions of the
Einstein-Maxwell-Dilaton gravity model (EMD) in $5$-dimensions, the action of
which being%
\[
I =\frac{1}{16\pi G}\int d^{5}x\sqrt{g}\left(  R-\frac{1}{2}\left(
\partial \phi \right)  ^{2}-\frac{1}{4}e^{-\alpha \phi}F^{2}\right) .
\]
Except that these solutions have no supersymmetry, they carry every typical
properties of all the black ring solutions known to this date, e.g. they all
have horizon(s) of $S^{2}\times S^{1}$ topology, they all have $3$ Killing
coordinates which determine their local symmetries, they must either be
rotating along the $S^{1}$ or carry electronic charge or local dipole charge
to prevent them from collapsing into black holes with horizons of $S^{3}$
topology, etc. For historical reasons the black ring solutions are all
formulated in terms of $C$-metric-like \cite{Dowker:1993bt} coordinates, i.e. the $5$ spacetime
coordinates are chosen such that%
\[
x^{\mu}=(t,y,\psi,x,\varphi),
\]
where $t$ is the coordinate time, $-\infty<y\leq-1$ represent a radial
coordinate, $-1\leq x\leq1$ and $\psi,\varphi$ are $3$ different angular
coordinates, and to avoid conical singularities, the angular coordinates
$\psi$ and $\varphi$ respectively must be given certain unusual periods. The
$3$ Killing coordinates are respectively $t,\psi$ and $\varphi$, however the
vector $\partial_{t}$ does not necessarily be always timelike (when it
doesn't, it implies the existence of an ergo-region).

For later use, let us now outline each black ring solutions of the EMD theory
explicitly. They are

\begin{itemize}
\item Neutral black ring (first obtained R. Emparan in \cite{hep-th/0402149}%
):
\begin{eqnarray*}
ds^{2}  &  =&-\frac{F(y)}{F(x)}\left(  dt+C(\nu,\lambda)R\frac{1+y}{F(y)}%
d\psi \right)  ^{2}\\
&  &+\frac{R^{2}}{(x-y)^{2}}F(x)\left[  -\frac{G(y)}{F(y)}d\psi^{2}%
-\frac{dy^{2}}{G(y)}+\frac{dx^{2}}{G(x)}+\frac{G(x)}{F(x)}d\varphi^{2}\right]
,
\end{eqnarray*}
where%
\begin{eqnarray*}
F(\xi)  &  =&1+\lambda \xi,\quad G(\xi)=(1-\xi^{2})(1+\nu \xi),\\
C(\nu,\lambda)  &  =&\sqrt{\lambda(\lambda-\nu)\frac{1+\lambda}{1-\lambda}},
\end{eqnarray*}
In order to avoid conical singularities at $x=-1$ and $y=-1$, the angular
coordinates $\varphi$ and $\psi$ are chosen to have the periodicity%
\[
\Delta \varphi=\Delta \psi=2\pi \frac{\sqrt{1-\lambda}}{1-\nu}.
\]
The parameters $\lambda,\nu$ are dimensionless and take values in the range $
0<\nu \leq \lambda<1$, and to remove the conical singularity also at $x=1$, $\lambda$ and $\nu$ must
be related to each other via%
\[
\lambda=\frac{2\nu}{1+\nu^{2}}.
\]
The metric has a killing horizon at $y=y_H=-1/\nu$, where the coefficient of $dy^{2}$ diverges.

\item Dipole black ring (also found in \cite{hep-th/0402149}):
\begin{eqnarray*}
ds^{2}  &  =&-\frac{F(y)}{F(x)}\left(  \frac{H(x)}{H(y)}\right)  ^{N/3}\left(
dt+C(\nu,\lambda)R\frac{1+y}{F(y)}d\psi \right)  ^{2}\\
 & & +\frac{R^{2}}{(x-y)^{2}}F(x)
\left(  H(x)H(y)^{2}\right)  ^{N/3}\\
&  &\times \left[
-\frac{G(y)}{F(y)H(y)^{N}}d\psi^{2}-\frac{dy^{2}}{G(y)}+\frac{dx^{2%
}}{G(x)}+\frac{G(x)}{F(x)H(x)^{N}}d\varphi^{2}\right]  ,
\end{eqnarray*}
where%
\begin{eqnarray*}
H(\xi) &=&1-\mu \xi,\\
0 &\leq& \mu<1,\\
\alpha^{2} &=&\frac{4}{N}-\frac{4}{3},
\end{eqnarray*}
This metric accompanied by the following electromagnetic one form potential and dilaton
field:%
\begin{eqnarray*}
\mathcal{A}  &  =&\left(C(\nu,-\mu)\sqrt{N}R\frac{1+x}{H(x)}+k\right) d\varphi,\\
e^{-\phi}  &  =&\left(  \frac{H(x)}{H(y)}\right)  ^{N\alpha/2}.
\end{eqnarray*}
The case $N=1$, or $\alpha=\sqrt{8/3}$, corresponds to Kaluza-Klein coupling.

The condition for removing conical singularities at $x=-1$ and $y=-1$ gives
rise to periods for the coordinates $\varphi$ and $\psi$:%
\[
\Delta \varphi=\Delta \psi=4\pi \frac{\sqrt{F(-1)H(-1)^{N}}}{G^{\prime}(-1)}%
=2\pi \frac{\sqrt{(1-\lambda)(1+\mu)^{N}}}{1-\nu},
\]
while the elimination of conical singularity at $x=+1$ requires
\[
\Delta \varphi=4\pi \frac{\sqrt{F(+1)H(+1)^{N}}}{G^{\prime}(+1)}%
=2\pi \frac{\sqrt{(1+\lambda)(1-\mu)^{N}}}{1+\nu}.
\]
The horizon is also located at $y=y_H=-1/\nu$. At $\mu=0$, this solution degenerates into the neutral
black ring just mentioned.

\item Electronically charged black ring: (found in \cite{hep-th/0412153})
\begin{eqnarray*}
ds^{2}  & =&-V_{k}(x,y)^{-2N/3}\frac{F(x)}{F(y)}dt^{2}+V_{k}(x,y)^{N/3}%
\frac{R^{2}}{(x-y)^{2}}\\
& &\times \left[  -F(x)\left(  (1-y^{2})d\psi^{2}+\frac{F(y)}{1-y^{2}}%
dy^{2}\right)  +F(y)^{2}\left(  \frac{dx^{2}}{1-x^{2}}+\frac{1-x^{2}}%
{F(x)}d\varphi^{2}\right)  \right]  ,\\
\mathcal{A}  & =&V_{k}(x,y)^{-1}k\frac{1-\frac{F(x)}{F(y)}}{1-k^{2}}dt,\\
e^{-\phi}  & =&V_{k}(x,y)^{\alpha/2},
\end{eqnarray*}
where $0<\lambda<1$ and
\begin{eqnarray*}
F(\xi)  & =&1-\lambda \xi,\\
V_{k}(x,y)  & =&\frac{1-k^{2}\frac{F(x)}{F(y)}}{1-k^{2}}.
\end{eqnarray*}

The horizon is located at $y=y_{H}=-\infty$. To avoid conical singularities
at $x=-1$ and $y=-1$, the coordinates $\varphi$
and $\psi$ must be chosen to have the periods%
\[
\Delta \varphi=\Delta \psi=2\pi \sqrt{1+\lambda}.
\]
At $x=+1$, then, there will be inevitably a conical singularity. Alternatively
we may demand regularity at $x=+1$ by requiring the period for $\varphi$ to be%
\[
\Delta \varphi=2\pi \sqrt{1-\lambda},
\]
but then at $x=-1$ a conical singularity arises. So, conical singularities always exist in this spacetime.

\item Dyonic black ring (only known at KK coupling \cite{hep-th/0412153}):
\begin{eqnarray*}
ds^{2}  & =&-\frac{F(x)}{F(y)}\frac{(dt+R\sqrt{\lambda \nu}(1+y)d\psi)^{2}%
}{V_{k}(x,y)^{2/3}}\\
& +&\frac{R^{2}}{(x-y)^{2}}V_{k}(x,y)^{1/3}\left[  -F(x)\left(  G(y)d\psi
^{2}+\frac{F(y)}{G(y)}dy^{2}\right)  + F(y)^{2}\left(  \frac{dx^{2}}{G(x)}+\frac{G(x)}{F(x)}d\varphi
^{2}\right)  \right]  ,\\
\mathcal{A}  & =&V_{k}(x,y)^{-1}k\left[  \frac{1-\frac{F(x)}{F(y)}}{1-k^{2}}%
dt-R\sqrt{\lambda \nu}(1+y)\frac{F(x)}{F(y)}d\psi \right]  ,\\
e^{-\phi}  & =&V_{k}(x,y)^{\sqrt{2/3}},
\end{eqnarray*}
where $G(\xi)=(1-\xi^{2})(1-\nu(1-k^{2})\xi)$. At $\nu=0$, this solution
degenerates into the static electronically charged black ring solution.

The condition to remove conical singularities at $x=-1$ and $y=-1$ reads
\[
\Delta \varphi=\Delta \psi=2\pi \sqrt{\frac{1+\lambda}{1+\nu(1-k^{2})}}.
\]
Similar requirement at $x=+1$ yields
\[
\Delta \varphi=2\pi \sqrt{\frac{1-\lambda}{1-\nu(1-k^{2})}},
\]
so one is led to the condition%
\[
\lambda=\nu(1-k^{2}).
\]
One sees that this is impossible if $\nu=0$ but $\lambda \neq0$. Moreover, the horizon located at
\[
y=y_{H}=\frac{1}{\nu(1-k^{2})}
\]
does not degenerate to the one for the electronically charged black ring in the limit $\nu=0$.
\end{itemize}

In the next section, we shall study the Parikh-Wilczek tunnelling from the
above black ring spacetimes. For this purpose, the original form of the
metrics are not the most pertinent ones. Actually, it is the ADM form \cite{ADM} of the
metrics which is most appropriate for our purpose. From the explicit form of
the metrics list above, it can be seen that they all can be casted into the
following form,
\begin{equation}
\label{adm}ds^{2}=-Adt^{2}+B^{-1}dy^{2}+g_{\psi \psi}(d\psi+N^{\psi}%
dt)^{2}+g_{xx}dx^{2}+g_{\varphi \varphi}d\varphi^{2},
\end{equation}
where of course the coefficient functions $A,B,g_{xx},g_{\psi \psi}%
,g_{\varphi \varphi}$ and $N^{\psi}$ are all functions of $x,y$ only and for
each of the above metrics they have their own different values. Moreover, the
coefficients in the ADM metrics obey the following unique
property
\[
A(x,y_{H}) = B(x,y_{H})=0, \qquad N^{\psi}(x,y_{H})= \Omega_{H},
\]
where $y_{H}$ represents the value of $y$ at the event horizon and $\Omega
_{H}$ is just the angular velocity of the black ring at the event horizon
\footnote{Notice that the angular velocity at the horizon given in
\cite{hep-th/0402149} differs from the $\Omega_{H}$ here by a factor of
$\frac{2\pi}{\Delta \psi}$. This is because the angular coordinate $\psi$ has
the unusual period $\Delta \psi$ in order to avoid conical singularities, but
the angular velocity in \cite{hep-th/0402149} is measured by a rescaled angle
$\psi^{\prime}$ which has normal period $2\pi$. Here we prefer to use the
original coordinate $\psi$ to define angular velocity, because it is the time
derivative of this variable which appears in the expression for the null
Killing vector $\xi= k +\Omega_{H} m = \frac{\partial}{\partial t} +
\Omega_{H} \frac{\partial}{\partial \psi}$.}. The electromagnetic one form
potentials associated with the above black ring solutions also
have some common properties. For instance, the components $\mathcal{A}_{y}$ and $\mathcal{A}_{x}$
always vanish,
\begin{equation}
\label{potential}\mathcal{A}_{\mu}= (\mathcal{A}_{t}, 0, \mathcal{A}_{\psi}, 0, \mathcal{A}_{\varphi}),
\end{equation}
where $\mathcal{A}_{t} = - \Phi$ is the scalar potential of the electro-magnetic field.
These facts will be very useful when we study the tunnelling from the black rings.

\section{The tunnelling method and its application to the black ring
spacetimes}

\subsection{The tunnelling method}

The tunnelling approach is based on the following simple idea.
Assume a (neutral) relativistic particle is emitted (via a
virtual process) from inside the horizon (of a static, non-rotating black
hole) to infinity. In this process the particle is travelling backward in
time, so the action of the particle must involve an imaginary part which
represents the tunnelling amplitude, i.e.
\[
\mathrm{Amplitude}\propto e^{i I}.
\]
Accordingly, the emission rate obeys the relation%
\[
\Gamma \propto|\mathrm{Amplitude}|^{2}\propto e^{-2\mathrm{Im} I}.
\]
This is further identified with the Boltzmann weight for the radiation
\[
\Gamma \propto e^{-\beta \mathcal{E}},
\]
with $\mathcal{E}$ being the energy of the emitted particle.
In this consideration the back reaction of the emitted particle on the black hole metric is
ignored and hence the radiation spectrum is thermo. The Hawking temperature
of the black hole is easy to be read off as $T=\frac{1}{\beta}$. A
more careful analysis must involve the back reaction of the emitted particle
on the black hole metric, which modifies the dependence of the
exponent of $\Gamma$ on the particle's energy $\mathcal{E}$, which in general becomes nonlinear
(and hence non-thermo spectrum). So the key to the tunnelling approach is the evaluation of the imaginary
part of the action of the virtual particle.

The simplest example of tunnelling process is the case of Schwarzchild black hole
studied in Parikh and Wilczek's work \cite{hep-th/9907001}. The metric in a
non-singular coordinate (which was first obtained by Painlev\'{e} \cite{Isreal} and brought
back to the modern audience by Kraus and Wilczek in \cite{gr-qc/9406042}) reads
\begin{equation}\label{nonsing}
ds^{2}=-\left( 1-\frac{2M}{r}\right) dt^{2}+2\sqrt{\frac{2M}{r}}dt dr+dr^{2} +
r^{2} d\Omega_{2}^{2}.
\end{equation}
The radial null geodesics obey the following equation:
\[
\frac{dr}{dt}=\pm1-\sqrt{\frac{2M}{r}}.
\]
Emission of a particle of energy $\omega$ results in a back reaction
which effectively sees the metric as one with black hole mass $M-\omega$. So
the imaginary part of the action is
\begin{eqnarray*}
\mathrm{Im} I & =&\mathrm{Im} \int_{r_{\rm in}}^{r_{\rm out}} p_{r} dr =\mathrm{Im}
\int_{r_{\rm in}}^{r_{\rm out}} \int_{0}^{p_{r}} dp_{r}^{\prime}dr\\
& =&\mathrm{Im} \int_{M}^{M-\omega}\int_{r_{\rm in}}^{r_{\rm out}} \frac{dr}{\dot{r}}dH
=-\mathrm{Im} \int_{0}^{\omega}\int_{r_{\rm in}}^{r_{\rm out}} \frac{dr}{1-\sqrt
{\frac{2(M-\omega)}{r}}}d\omega \\
& =&4\pi \omega \left( M-\frac{\omega}{2}\right) .
\end{eqnarray*}
Therefore, the emission rate is given by
\begin{equation}\label{PW}
\Gamma \propto e^{-2\mathrm{Im}I}=e^{\Delta S_{BH}},
\end{equation}
where $\Delta S_{BH}$ is the difference of the Bekenstein-Hawking entropy of the hole after and before the
tunnelling process. This relation is exactly the Keski-Vakkuri-Kraus relation \cite{hep-th/9610045}
mentioned in the introduction, which is repeatedly found to hold in all black hole spacetimes.

From this example one sees that the direct application of Parikh-Wilczek's
approach to black ring metrics has some difficulties, e.g. it is cumbersome to find a
non-singular coordinate as in (\ref{nonsing}); it is difficult to determine the pure radial geodesics
because of non-spherical symmetries; it is difficult to determine the effect of back reaction, which is needed to
make the integration over $\omega$ and it is difficult to consider the lose of angular momentum and/or electronic
discharge, etc. Some of these difficulties may be overcome after some elaborate works.
However it is more advantageous to address all the difficulties at once by
introducing a unified covariant framework for studying the tunnelling process. Fortunately,
such a covariant framework already exist for the cases of rotating black holes, i.e. the ANVZ approach
\cite{hep-th/0503081}. In the next subsection we will slightly adapt the ANVZ approach
to the charged rotating cases which we then apply to the black ring solutions
listed in section 2. The effect of back reactions will also be taken into account following \cite{gr-qc/0504113}.

\subsection{ANVZ approach for charged rotating black rings}

The most efficient way to find the imaginary part of the emitted particle is
to solve the particle's relativistic Hamilton-Jacobi equation
\begin{equation}
\label{HJ}g^{\mu \nu}(\partial_{\mu} I -\mathcal{Q} A_{\mu}) (\partial_{\nu} I
- \mathcal{Q} A_{\nu}) +m^{2}=0,
\end{equation}
where $\mathcal{Q}$ is the electronic charge of the particle and $m$ is its
stationary mass. Before solving this equation, let us stress again that what
is relevant here is only the \emph{imaginary part} of the action, which comes
from the contribution of the particle moving from somewhere very near but
inside the horizon ($y=y_{\mathrm{in}}=y_{H}-0^{+}$) to somewhere very near
but outside the horizon ($y=y_{\mathrm{out}}=y_{H} +0^{+} $). Therefore, we
need only to consider the local behavior of the metrics near the horizon.
Under such assumptions, we may make a local change of coordinate
\[
\psi \rightarrow \chi= \psi+ \Omega_{H} t
\]
and keep the other coordinates unchanged, so that after the transformation,
the ADM form of the metric (\ref{adm}) behaves locally like a diagonal
metric,
\[
ds^{2}=-Adt^{2}+B^{-1}dy^{2}+g_{\psi \psi}d\chi^{2}+g_{xx}dx^{2}+g_{\varphi
\varphi}d\varphi^{2}.
\]
Note that the above metric is exact only at the horizon. However, we shall see
that this is the only point we shall be dealing with, so the above
approximation actually gives rise to \emph{exact results} for the tunnelling
amplitude and the Hawking temperature.

Now let us make the following ansatz for the solution $I$,
\begin{eqnarray} \label{ansatz}
I =-\mathcal{E}t+\mathcal{J}\chi+\mathcal{L}\varphi+W(x,y),
\end{eqnarray}
where $\mathcal{E}$, $\mathcal{J}$ and $\mathcal{L}$ are all real constants
which represents the energy and angular momenta with respect to the angles
$\chi$ and $\varphi$ respectively. It is clear that changing back into the
coordinate $\psi$ would only effectively change $\mathcal{E}$ into
$\mathcal{E}-\Omega_{H} \mathcal{J}$, i.e.
\begin{eqnarray*}
I =-(\mathcal{E-}\Omega_{H} \mathcal{J})t+\mathcal{J}\psi+\mathcal{L}%
\varphi+W(x,y).
\end{eqnarray*}
Inserting this into (\ref{HJ}), we get
\begin{eqnarray*}
\frac{\partial}{\partial y}W(x,y)  & \simeq &\frac{1}{\sqrt{AB}}\sqrt{\left(
\mathcal{E-}\Omega_{H}\mathcal{J}+\mathcal{Q} \mathcal{A}_{t} \right)  ^{2}-A\left(
m^{2}+\Psi(x,y)\right) },\\
\Psi(x,y) & \equiv& \left(  g_{xx}\right)  ^{-1}W_{,\,x}(x,y)^{2}+\left(
g_{\varphi \varphi}\right)  ^{-1}(\mathcal{L}-\mathcal{Q} \mathcal{A}_{\varphi}%
)^{2}+\left(  g_{\psi \psi}\right)  ^{-1}(\mathcal{J}-\mathcal{Q} \mathcal{A}_{\psi}%
)^{2},
\end{eqnarray*}
where $\simeq$ stands for ``equals near the horizon''. Now limiting to the
$s$-wave contribution, i.e. suppressing the angular contribution
$\Psi(x,y)$, the imaginary part of the action can be written as
\begin{eqnarray*}
\mathrm{Im} I  & =& \mathrm{Im}W(x,y)\simeq\mathrm{Im}\int_{y_{\mathrm{in}}%
}^{y_{\mathrm{out}}} dy\left. \frac{\partial}{\partial y}W(x,y) \right|
_{\Psi(x,y) \rightarrow0}\\
& =&\mathrm{Im}\int_{y_{\mathrm{in}}}^{y_{\mathrm{out}}} dy \frac{1}{\sqrt{AB}%
}\sqrt{\left(  \mathcal{E-}\Omega_{H}\mathcal{J} + \mathcal{Q} \mathcal{A}_{t} \right)
^{2}-A m^{2}}.
\end{eqnarray*}
The imaginary part of $W(x,y)$ arises from the pole contribution of the factor
$\frac{1}{\sqrt{AB}}$ at $y=y_{H}$. However, direct evaluation of the pole
contribution is highly coordinate dependent because of lack of covariance. The
correct way to make the integration is to make use of the \emph{proper spacial
distance} (this is where ANVZ make their important observation)
\[
d\sigma^{2}=B^{-1}(x,y)dy^{2}+g_{\psi \psi}(d\psi+\Omega_{H}dt)^{2}%
+g_{xx}dx^{2}+g_{\varphi \varphi}d\varphi^{2}.
\]
Under the $s$-wave approximation, we have
\[
\sigma= \int \frac{dy}{\sqrt{B(x,y)}}.
\]
In the near horizon limit, we get
\[
\sigma=\frac{2}{\sqrt{B_{,\,y}(x,y_{H})}}\sqrt{y-y_{H}}+...
\]
\[
\frac{1}{\sqrt{A(x,y)}}=\frac{2}{\sqrt{A_{,\,y}(x,y_{H})B_{,\,y}(x,y_{H})}%
}\frac{1}{\sigma}+...
\]
Therefore,
\begin{eqnarray*}
\mathrm{Im} I  & =&\frac{2}{\sqrt{A_{,\,y}(x,y_{H})B_{,\,y}(x,y_{H})}}
\mathrm{Im}\int d\sigma \frac{1}{\sigma}\left(  \mathcal{E-}\Omega
_{H}\mathcal{J}+ \mathcal{Q} \mathcal{A}_{t}\right) .
\end{eqnarray*}
The integration on the RHS must be evaluated along a virtual path starting
from a small negative value of $\sigma$ to some positive value of $\sigma$.
Thus the pole contribution from the integrand $\frac{1}{\sigma}$ is imaginary
and can be obtained by slightly deforming the integration contour from the
real $\sigma$-axis to the lower complex $\sigma$-plane which avoids the pole
$\sigma=0$ counterclockwise. In the end,
\begin{eqnarray}
\mathrm{Im} I  & =&\frac{2}{\sqrt{A_{,\,y}(x,y_{H})B_{,\,y}(x,y_{H})}}
\times \frac{1}{2}\mathrm{res}|_{\sigma=0}\left( \frac{\mathcal{E-}\Omega
_{H}\mathcal{J}+ \mathcal{Q} \mathcal{A}_{t}}{\sigma}\right) \nonumber\\
& =&\frac{2\pi \left(  \mathcal{E}-\Omega_{H}\mathcal{J}- \Phi_{H}\mathcal{Q}
\right) }{\sqrt{A_{,\,y}(x,y_{H})B_{,\,y}(x,y_{H})}}. \label{ImI}
\end{eqnarray}
Note that the above is only the contribution from a single emission particle
and the back reaction of the particle on the black ring metric is not taken
into account. For accumulated emission of energy $\mathcal{E}$ and angular momentum
$\mathcal{J}$, one changes the last formula into \cite{gr-qc/0504113}
\[
\mathrm{Im} I =\frac{2\pi}{\sqrt{A_{,\,y}(x,y_{H})B_{,\,y}(x,y_{H})}}%
\int \left(  d\mathcal{E}-\Omega_{H} d\mathcal{J} - \Phi_{H} d\mathcal{Q}%
\right).
\]
The emission rate is then
\[
\Gamma \propto \exp \left(  -2\mathrm{Im}I \right)  =\exp \left(  -\beta \int(
d\mathcal{E}-\Omega_{H} d\mathcal{J}- \Phi_{H} d\mathcal{Q})\right) .
\]
From the last formula one reads out the expression for the inverse temperature:
\begin{equation}
\label{temperature}\beta=\frac{4\pi}{\sqrt{A_{,\,y}(x,y_{H})B_{,\,y}(x,y_{H}%
)}}=\frac{1}{T}.
\end{equation}

Note that there are diverse proofs and/or verifications for the correctness of
the first law of thermodynamics for black rings \cite{hep-th/0402149, hep-th/0509144, hep-th/0509150}, i.e.
\[
dE = T dS_{BH} + \Omega_{H} dJ + \Phi_{H} dQ \Rightarrow dS_{BH} =\frac{dE
-\Omega_{H} dJ - \Phi_{H} dQ}{T},
\]
where $dE = -d\mathcal{E}, dJ = -d\mathcal{J}, dQ = -d\mathcal{Q}$, which are required by the conservation
of mass, angular momentum and electronic charge. So, the
emission rate can also be expressed as
\[
\Gamma \propto e^{\int dS_{BH}}= e^{\Delta S_{BH}},
\]
which is exactly the Keski-Vakkuri-Kraus relation first obtained in \cite{hep-th/9610045}. The present result
add some more evidence on the universality of this formula. Note that the effect of
electronic discharge as well as the back reaction of the emitted particles are
both included in the above considerations.

\subsection{Explicit results for black rings}

The construction made in the last subsection has led to a very simple formula
(\ref{temperature}) for calculating the Hawking temperature of black rings.
Now we apply this formula to the black ring metrics listed in section 2 to get
the explicit results.

\begin{itemize}
\item Neutral black ring:
\begin{eqnarray*}
A(x,y) & =&\frac{F(y)}{F(x)}\left(  1-\frac{C(\nu,\lambda)^{2}(1+y)^{2}}%
{\frac{1}{(x-y)^{2}}F(x)^{2}G(y)+C(\nu,\lambda)^{2}(1+y)^{2}}\right) ,\\
B(x,y) & =&-\left( \frac{R^{2}}{(x-y)^{2}}\frac{F(x)}{G(y)}\right) ^{-1},
\end{eqnarray*}
\[
\Rightarrow T=\frac{1}{4\pi R}\frac{1+\nu}{\nu^{1/2}}\sqrt{\frac{1-\lambda
}{\lambda(1+\lambda)}}.
\]

\item Dipole black ring:
\begin{eqnarray*}
A(x,y)  &  =&\frac{F(y)}{F(x)}\left(  \frac{H(x)}{H(y)}\right)  ^{N/3}\left(
1-\frac{C(\nu,\lambda)^{2}(1+y)^{2}}{\frac{1}{(x-y)^{2}}F(x)^{2}%
G(y)+C(\nu,\lambda)^{2}(1+y)^{2}}\right)  ,\\
B(x,y)  &  =&-\left(  \frac{R^{2}}{(x-y)^{2}}\left(  H(x)H(y)^{2}\right)
^{N/3}\frac{F(x)}{G(y)}\right)  ^{-1};
\end{eqnarray*}

\[
\Rightarrow T=\frac{1}{4\pi R}\frac{\nu^{(N-1)/2}(1+\nu)}{(\mu+\nu)^{N/2}%
}\sqrt{\frac{1-\lambda}{\lambda(1+\lambda)}}.
\]

\item electronically charged:
\[
T=\frac{1}{4\pi R\lambda}(1-k^{2})^{1/2},
\]

\item dyonic:
\[
T=\frac{1-\nu(1-k^{2})}{4\pi R}\sqrt{\frac{1-k^{2}}{\lambda(\lambda-\nu(1-k^{2}))}}.
\]

\end{itemize}

All these results agree exactly with the surface gravity results presented in
the original papers \cite{hep-th/0402149, hep-th/0412153}. For the dipole black ring the same Hawking
temperature has also been obtained in \cite{hep-th/0509144} using the so-called quasilocal formalism.

\subsection{Hawking radiation at the Kaluza-Klein coupling}

Before ending the present section, let us make some further remarks on the black ring radiation at the Kaluza-Klein
coupling $\alpha=\sqrt{8/3}$. In such cases, the tunnelling process can also be studied in the framework of
$6$-dimensional pure Einstein gravity theory. The Hamilton-Jacobi
equation for the emitted particle can be written as%
\begin{equation}\label{HJ6}
G^{MN}\partial_{M}I\partial_{N}I+m^{2}=0,
\end{equation}
where $G_{MN}$ is given by the $6$-dimensional Kaluza-Klein metric%
\begin{equation}\label{6d}
ds_{6}^{2}=e^{\sqrt{1/6}\phi}ds_{5}^{2}+e^{-\sqrt{3/2}\phi}(dz+\mathcal{A}%
)^{2},
\end{equation}
where $ds_{5}^{2}$ corresponds to the $5$-dimensional black ring metric,
$\mathcal{A}=A_{\mu}dx^{\mu}$ is the electro-magnetic one form potential and
$z$ is the $6$-th spacetime coordinate.

Now the ansatz (\ref{ansatz}) for the solution is (locally) changed into ($\zeta \equiv z - \Phi_H t$)
\begin{eqnarray}
I &=&-\mathcal{E}t+\mathcal{J}\chi+\mathcal{L}\varphi-\mathcal{Q}\zeta+ W(x,y) \nonumber\\
&=&-(\mathcal{E}-\Omega_H \mathcal{J} - \Phi_H \mathcal{Q})t + \mathcal{J}\psi+\mathcal{L}\varphi-\mathcal{Q}z+ W(x,y).
\label{ansatz6}
\end{eqnarray}
It is crucial to point out that in the $6$-d metric (\ref{6d}), if $ds_5^2$ is written in the ADM form,
then $ds_6^2$ is automatically an ADM metric, and the coefficient functions $A(x,y)$ and $B(x,y)$ undergo the
following transformation:
\begin{eqnarray*}
&A(x,y)\rightarrow e^{\sqrt{1/6}\phi(x,y)}A(x,y),\quad
B(x,y) \rightarrow e^{-\sqrt{1/6}\phi(x,y)}B(x,y).&
\end{eqnarray*}
Therefore, inserting the ansatz (\ref{ansatz6}) into the Hamilton-Jacobi equation (\ref{HJ6}) gives rise to the
same solution (\ref{ImI}) as in the $5$-dimensional cases because the product $A(x,y)B(x,y)$ is kept unchanged.
Notice that $\mathcal{Q}$ here is to be understood as the angular momentum which generates rotations along
the $z$ direction. That the imaginary part of the emitted particle (and hence the Hawking temperature) is the
same from both the $5$-dimensional and $6$-dimensional perspectives is not difficult to understand. Actually, since the
Kaluza-Klein metric (\ref{6d}) contains only the zero KK mode, the tunnelling particles only escape from the black rings and
run into the $5$-dimensional sub-spacetime described by $ds_5^2$. If there were nonzero KK modes involved, the Hawking
temperatures of from the $5$-d and $6$-d perspectives would have been different.

\section{Discussions and concluding remarks}

In this paper, we studied the problem of Hawking radiation from black ring solutions
of the $5$-dimensional EMD gravity theory. It is shown that provided the first law of thermodynamics for
black rings is correct, the emission rate is related to the loss of Bekenstein-Hawking entropy of the black rings
during the radiation. This result is in agreement with earlier studies of Hawking radiation in terms of
tunnelling for other types of black holes and it seems to be a universal property for the tunnelling
process. We also obtained a simple formula for calculating black ring temperatures, which leads to exactly the same result
as compared to the classical surface gravity method and the quasilocal formalism. The method we used can be applied
straightforwardly to the cases of other black rings/holes.

Since black rings are a new class of black spacetime solutions,
it seems interesting to make further analysis on their various properties, especially the result of microstate counting
using string theory methods is still to be compared with (an analysis on microstate counting for the dipole black ring
to leading order is already made in \cite{hep-th/0402149}). Also, the stability analysis,
both from dynamical and thermodynamical perspectives, for the black rings is also an interesting field of further study.

\section*{Acknowledgements}
The author would like to thank X.H. Meng for discussions. This work is supported by the National Natural Science Foundation
of China through grant No.90403014.


\begin{thebibliography}{10}

\bibitem{Hawking:1974sw}
S.~W. Hawking, ``Particle creation by black holes,'' {\em Commun. Math. Phys.}
  {\bf 43} (1975)
199--220.

\bibitem{gr-qc/9408003}
P.~Kraus and F.~Wilczek, ``Selfinteraction correction to black hole radiance,''
  {\em Nucl. Phys.} {\bf B433} (1995) 403--420,
\href{http://www.arXiv.org/abs/gr-qc/9408003}{{\tt gr-qc/9408003}}.

\bibitem{hep-th/9411219}
P.~Kraus and F.~Wilczek, ``Effect of selfinteraction on charged black hole
  radiance,'' {\em Nucl. Phys.} {\bf B437} (1995) 231--242,
\href{http://www.arXiv.org/abs/hep-th/9411219}{{\tt hep-th/9411219}}.

\bibitem{hep-th/9610045}
E.~Keski-Vakkuri and P.~Kraus, ``Microcanonical D-branes and back reaction,''
  {\em Nucl. Phys.} {\bf B491} (1997) 249--262,
\href{http://www.arXiv.org/abs/hep-th/9610045}{{\tt hep-th/9610045}}.

\bibitem{hep-th/9907001}
M.~K. Parikh and F.~Wilczek, ``Hawking Radiation as Tunneling,'' {\em
  Phys.Rev.Lett.} {\bf 85} (2000) 5042--5045,
  \href{http://www.arXiv.org/abs/hep-th/9907001}{{\tt hep-th/9907001}}.

\bibitem{hep-th/0204107}
M.~K. Parikh, ``New coordinates for de Sitter space and de Sitter radiation,''
  {\em Phys. Lett.} {\bf B546} (2002) 189--195.

\bibitem{hep-th/0405160}
M.~K. Parikh, ``A secret tunnel through the horizon,'' {\em Int. J. Mod. Phys.}
  {\bf D13} (2004) 2351--2354,
\href{http://www.arXiv.org/abs/hep-th/0405160}{{\tt hep-th/0405160}}.

\bibitem{hep-th/0012134}
E.~C. Vagenas, ``Are extremal 2-D black holes really frozen?,'' {\em Phys.
  Lett.} {\bf B503} (2001) 399--403,
\href{http://www.arXiv.org/abs/hep-th/0012134}{{\tt hep-th/0012134}}.

\bibitem{hep-th/0108147}
E.~C. Vagenas, ``BTZ black holes and Hawking radiation,'' {\em Mod. Phys.
  Lett.} {\bf A17} (2002) 609--618.

\bibitem{hep-th/0109108}
E.~C. Vagenas, ``Quantum corrections to the Bekenstein-Hawking entropy of the
  BTZ black hole via self-gravitation,'' {\em Phys. Lett.} {\bf B533} (2002)
  302--306.

\bibitem{hep-th/0110289}
A.~J.~M. Medved, ``Radiation via tunneling in the charged BTZ black hole,''
  {\em Class. Quant. Grav.} {\bf 19} (2002) 589--598.

\bibitem{hep-th/0207247}
A.~J.~M. Medved, ``Radiation via tunneling from a de Sitter cosmological
  horizon,'' {\em Phys. Rev.} {\bf D66} (2002) 124009.

\bibitem{hep-th/0209185}
E.~C. Vagenas, ``Generalization of the KKW analysis for black hole radiation,''
  {\em Phys. Lett.} {\bf B559} (2003) 65--73.

\bibitem{hep-th/0503081}
M.~Angheben, M.~Nadalini, L.~Vanzo, and S.~Zerbini, ``Hawking radiation as
  tunneling for extremal and rotating black holes,'' {\em JHEP} {\bf 05} (2005)
  014,
\href{http://www.arXiv.org/abs/hep-th/0503081}{{\tt hep-th/0503081}}.

\bibitem{gr-qc/0504113}
A.~J.~M. Medved and E.~C. Vagenas, ``On Hawking radiation as tunneling with
  back-reaction,'' {\em Mod. Phys. Lett.} {\bf A20} (2005) 2449--2454,
\href{http://www.arXiv.org/abs/gr-qc/0504113}{{\tt gr-qc/0504113}}.

\bibitem{gr-qc/0505015}
A.~J.~M. Medved and E.~C. Vagenas, ``On Hawking radiation as tunneling with
  logarithmic corrections,'' {\em Mod. Phys. Lett.} {\bf A20} (2005)
  1723--1728,
\href{http://www.arXiv.org/abs/gr-qc/0505015}{{\tt gr-qc/0505015}}.

\bibitem{Zhang:2005xt}
J.-Y. Zhang and Z.~Zhao, ``Hawking radiation of charged particles via tunneling
  from the Reissner-Nordstroem black hole,'' {\em JHEP} {\bf 10} (2005) 055.

\bibitem{hep-th/0511123}
Q.-Q. Jiang and S.-Q. Wu, ``Hawking radiation of charged particles as tunneling
  from Reissner-Nordstrom-de Sitter black holes with a global monopole,''
  \href{http://www.arXiv.org/abs/hep-th/0511123}{{\tt hep-th/0511123}}.

\bibitem{hep-th/0511250}
M.~Nadalini, L.~Vanzo, and S.~Zerbini, ``Hawking Radiation as Tunneling: the
  D-dimensional rotating case,''
  \href{http://www.arXiv.org/abs/hep-th/0511250}{{\tt hep-th/0511250}}.

\bibitem{gr-qc/0512153}
J.~Zhang and Z.~Zhao, ``Charged particles' tunnelling from the Kerr-Newman
  black hole,'' \href{http://www.arXiv.org/abs/gr-qc/0512153}{{\tt
  gr-qc/0512153}}.

\bibitem{hep-th/0512121}
J.~Zhang, Y.~Hu, and Z.~Zhao, ``Information loss in black hole evaporation,''
  \href{http://www.arXiv.org/abs/hep-th/0512121}{{\tt hep-th/0512121}}.

\bibitem{gr-qc/0601018}
Y.-p. Hu, J.-y. Zhang, and Z.~Zhao, ``The relation between Hawking radiation
  via tunnelling and the laws of black hole thermodynamics,''
  \href{http://www.arXiv.org/abs/gr-qc/0601018}{{\tt gr-qc/0601018}}.

\bibitem{hep-th/0505266}
M.~Arzano, A.~J.~M. Medved, and E.~C. Vagenas, ``Hawking radiation as tunneling
  through the quantum horizon,'' {\em JHEP} {\bf 09} (2005) 037,
\href{http://www.arXiv.org/abs/hep-th/0505266}{{\tt hep-th/0505266}}.

\bibitem{hep-th/0309092}
M.~R. Setare and E.~C. Vagenas, ``Self-gravitational corrections to the
  Cardy-Verlinde formula of Achucarro-Ortiz black hole,'' {\em Phys. Lett.}
  {\bf B584} (2004) 127--132,
\href{http://www.arXiv.org/abs/hep-th/0309092}{{\tt hep-th/0309092}}.

\bibitem{hep-th/0405186}
M.~R. Setare and E.~C. Vagenas, ``Self-gravitational corrections to the
  Cardy-Verlinde formula and the FRW brane cosmology in SdS(5) bulk,''
\href{http://www.arXiv.org/abs/hep-th/0405186}{{\tt hep-th/0405186}}.

\bibitem{hep-th/0402149}
R.~Emparan, ``Rotating circular strings, and infinite non-uniqueness of black
  rings,'' {\em JHEP} {\bf 03} (2004) 064,
\href{http://www.arXiv.org/abs/hep-th/0402149}{{\tt hep-th/0402149}}.

\bibitem{hep-th/0110258}
R.~Emparan and H.~S. Reall, ``Generalized Weyl solutions,'' {\em Phys. Rev.}
  {\bf D65} (2002) 084025.

\bibitem{hep-th/0110260}
R.~Emparan and H.~S. Reall, ``A rotating black ring in five dimensions,'' {\em
  Phys. Rev. Lett.} {\bf 88} (2002) 101101.

\bibitem{Dowker:1993bt}
F.~Dowker, J.~P. Gauntlett, D.~A. Kastor, and J.~H. Traschen, ``Pair creation
  of dilaton black holes,'' {\em Phys. Rev.} {\bf D49} (1994) 2909--2917.

\bibitem{hep-th/0412153}
H.~K. Kunduri and J.~Lucietti, ``Electrically charged dilatonic black rings,''
  {\em Phys. Lett.} {\bf B609} (2005) 143--149,
\href{http://www.arXiv.org/abs/hep-th/0412153}{{\tt hep-th/0412153}}.

\bibitem{ADM}
R.~Arnowitt, S.~Deser, and C.~Misner, {``The dynamics of general relativity''},
in {\em Gravitation: an introduction to current research}, pp227-265, John Wisley \& Sons, Inc. (1962),
Ed. by Louis Witten.

\bibitem{Isreal}
W.~Isreal,``Dark stars: the evolution of an idea", in {\em 300 years of Gravitation}, pp234,
Cambridge University Press (1987), Ed. by S. Hawking and W. Isreal.

\bibitem{gr-qc/9406042}
P.~Kraus and F.~Wilczek, ``A Simple stationary line element for the
  Schwarzschild Geometry, and some applications,''
\href{http://www.arXiv.org/abs/gr-qc/9406042}{{\tt gr-qc/9406042}}.

\bibitem{hep-th/0509144}
D.~Astefanesei and E.~Radu, ``Quasilocal formalism and black ring
  thermodynamics,''
\href{http://www.arXiv.org/abs/hep-th/0509144}{{\tt hep-th/0509144}}.

\bibitem{hep-th/0509150}
M.~Rogatko, ``Black rings and the physical process version of the first law of
  thermodynamics,'' {\em Phys. Rev.} {\bf D72} (2005) 074008,
\href{http://www.arXiv.org/abs/hep-th/0509150}{{\tt hep-th/0509150}}.

\end{thebibliography}

\providecommand{\href}[2]{#2}\begingroup\raggedright\endgroup

\end{document}